\documentclass[10pt]{article}
\usepackage{amsmath}
\usepackage{ictamxs}
\usepackage{graphicx}
\usepackage{epsfig}
\usepackage[small,bf]{subfigure}
\usepackage{times}

\input{tcilatex}

\begin{document}

%
%
%
%
\title{Linear waves and baroclinic instability in an inhomogeneous-density layered
primitive-equation ocean model}

\authors{\underline{F. J. Beron-Vera}\footnote{Electronic mail: \texttt{fberon@rsmas.miami.edu}.}, M. J. Olascoaga and J. Zavala-Garay}

\affiliations{RSMAS, University of Miami, Miami, FL 33149, USA}

\maketitle

\begin{summary}
We consider a multilayer generalization of Ripa's inhomogeneous-density single-layer
primitive-equation model. In addition to vary arbitrarily in horizontal
position and time, the horizontal velocity and buoyancy fields are allowed to
vary linearly with depth within each layer of the model. Preliminary results on linear
waves and baroclinic instability suggest that a configuration
involving a few layers may set the basis for a quite accurate
and numerically efficient ocean model.
\end{summary}%

\section{Introduction}
\vspace{-.1in}

Because of the physical and practical relevance of layered models, methods
to accommodate thermodynamic processes into these models have been
developed. The simplest method consists of allowing the buoyancy field to
vary in horizontal position and time, but keeping all dynamical fields as
depth independent. This is formally achieved by replacing the horizontal
pressure gradient by its vertical average \cite{Ripa-GAFD-93}. The resulting
inhomogeneous-density layered models, usually referred to as ``slab''
models, have been extensively used in ocean modeling.

Despite their widespread use, slab models are known to have several
limitations and deficiencies \cite{Ripa-DAO-99}. For instance, a slab
single-layer model cannot represent explicitly the thermal-wind balance
which dominates at low frequency. This balance is fundamental in processes
like cyclogenesis, which has been shown \cite{Eldevik-02} to be incorrectly
described with a slab single-layer model.

To cure the slab models limitations and deficiencies, Ripa \cite{Ripa-JFM-95}
proposed an improved closure to partially incorporate thermodynamic
processes in a one-layer model. In addition to allowing arbitrary velocity
and buoyancy variations in horizontal position and time, in Ripa's model the
horizontal velocity and buoyancy fields are allowed to vary linearly with
depth. The model, which has been generalized to an arbitrary number of
layers by Beron-Vera \cite{Beron-JFM-04}, enjoys a number of properties
which make it very promising for applications.

As a base test for the validity of the generalized Ripa's model we consider
linear waves, focusing particularly on vertical normal-mode phase speeds,
and classical baroclinic instability.

\section{The Generalized Model\label{model}}
\vspace{-.1in}

The starting point to generalize Ripa's single-layer model is to consider a
stack of $n$ active fluid layers of thickness $h_{i}(\mathbf{x},t)$. These
layers can be limited from below by an irregular, rigid surface and from
above by a soft surface or vice versa. We call the former possibility a
rigid bottom setting and the latter a rigid lid setting. Above (resp.,
below) the active fluid there is an inert, infinitely thick layer of lighter
(resp., denser) fluid in the rigid bottom (resp., lid) case. For the $i$%
th-layer horizontal velocity and buoyancy relative to the inert layer,
respectively, one must then write%
\begin{equation}
\mathbf{u}_{i}(\mathbf{x},\sigma ,t)=\mathbf{\bar{u}}_{i}(\mathbf{x}%
,t)+\sigma \mathbf{u}_{i}^{\sigma }(\mathbf{x},t),\quad \vartheta _{i}(%
\mathbf{x},\sigma ,t)=\bar{\vartheta}_{i}(\mathbf{x},t)+\sigma \vartheta
_{i}^{\sigma }(\mathbf{x},t).  \label{ansatz}
\end{equation}%
Here, the overbar stands for vertical average within the $i$th layer, and $%
\sigma $ is a scaled vertical coordinate which varies linearly from $\pm 1$
at the base of the $i$th layer to $\mp 1$ at the top of the $i$th layer.
[The upper (resp., lower) sign corresponds to the rigid bottom (resp., lid)
configuration.] The generalized model equations, which follow upon replacing
(\ref{ansatz}) in the continuously and arbitrarily stratified (i.e. exact)
primitive equations, namely rotating incompressible hydrostatic
Euler--Boussinesq equations, then take the form%
\begin{mathletters}\label{IL1n}
\begin{gather}
\partial _{t}h_{i}+\mathbf{\nabla }\cdot h_{i}\mathbf{\bar{u}}_{i}=0,
\tag{2a} \\
\overline{\mathrm{D}_{i}\vartheta _{i}}=0,\quad (\mathrm{D}_{i}\vartheta
_{i})^{\sigma }=0,  \tag{2b,c} \\
\overline{\mathrm{D}_{i}\mathbf{u}_{i}}+f\mathbf{\hat{z}}\times \mathbf{%
\bar{u}}_{i}+\overline{\mathbf{\nabla }p}_{i}=\mathbf{0},\quad (\mathrm{D}%
_{i}\mathbf{u}_{i})^{\sigma }+f\mathbf{\hat{z}}\times \mathbf{u}_{i}^{\sigma
}+(\mathbf{\nabla }p_{i})^{\sigma }=\mathbf{0}.  \tag{2d,e}
\end{gather}%
\end{mathletters}%
Here, $f$ is the Coriolis parameter (twice the local angular rotation
frequency); $\mathbf{\hat{z}}$ is the vertical unit vector; $\mathbf{\nabla }
$ is the horizontal gradient; $\overline{\mathrm{D}_{i}\mathbf{\bullet }}$
and $(\mathrm{D}_{i}\mathbf{\bullet })^{\sigma }$ are, respectively, the
vertical average and $\sigma $ components of the $i$th-layer material
derivative; and $\overline{\mathbf{\nabla }p}_{i}$ and $\left( \mathbf{%
\nabla }p_{i}\right) ^{\sigma }$ are, respectively, the vertical average and
$\sigma $ components of the $i$th-layer pressure gradient force. [For
details cf. Refs. 5,1.]

Some of the attractive properties of this model are the following. First,
the model can represent explicitly within each layer the thermal-wind
balance which dominates at low frequency. Second, volume, mass, buoyancy
variance, energy, and momentum are preserved by the dynamics. Third, because
of the possibility of vertical stratification within each layer,
thermodynamic processes (e.g. heat and freshwater inputs across the ocean
surface, vertical mixing, etc.) can be incorporated more realistically than
in slab models.

\section{Linear Waves and Baroclinic Instability\label{analysis}}
\vspace{-.1in}

\begin{figure}[t]
\centering{\subfigure[]{\epsfig{file=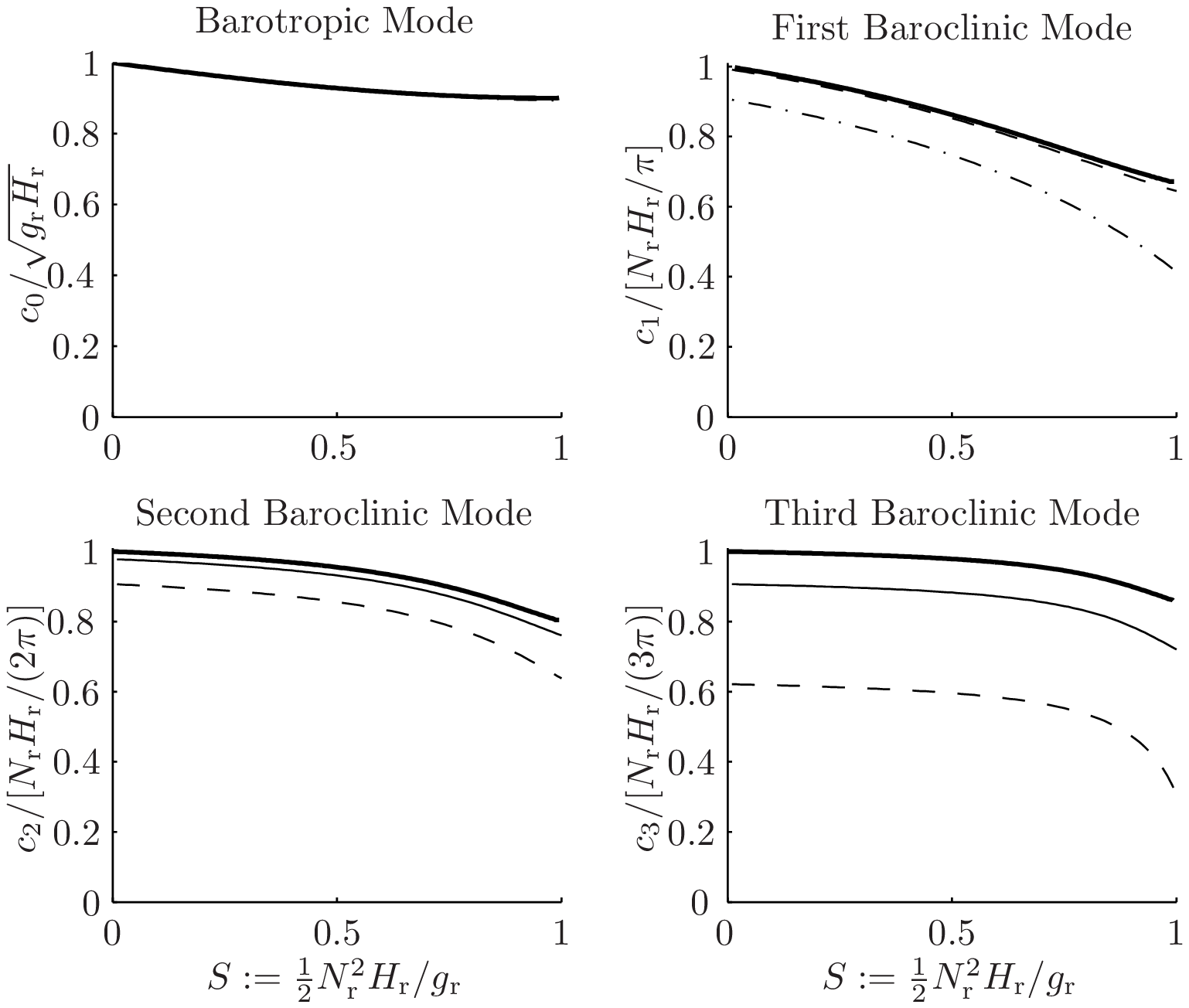,width=7.5cm}}\quad %
\subfigure[]{\epsfig{file=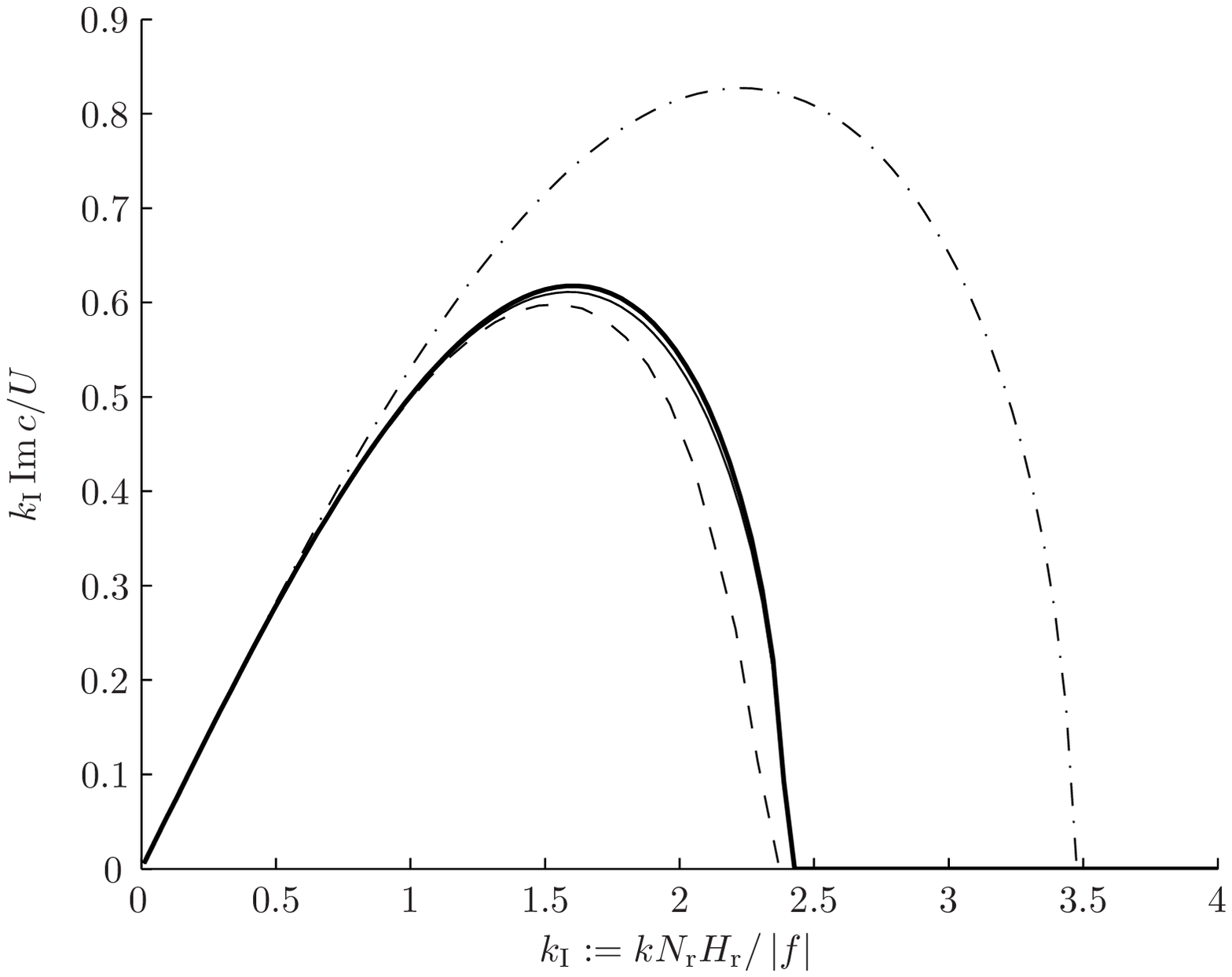,width=7.5cm}}} \caption{(a)
Phase speed of long gravity waves as a function of the
stratification in a reference state with no currents. (b) Growth
rate of the most unstable Eady wave as a function of wavenumber.
In both panels the exact result is indicated by a heavy-solid
line, and the layered model predictions with dot-dashed (one
layer), dashed (two layers), and light-solid (three layers)
lines.} \label{fig}
\end{figure}

System (\ref{IL1n}), linearized with respect to a \textit{reference state}
with no currents, can be shown to sustain the usual midlatitude and
equatorial gravity and vortical waves in several vertical normal modes. In
this work we concentrate on how well these modes are represented by
considering the phase speed of long gravity waves in a reference state
characterized by the stratification parameter $S:=\frac{1}{2}N_{\mathrm{r}%
}^{2}H_{\mathrm{r}}/g_{\mathrm{r}}.$ Here, $N_{\mathrm{r}}$ is the Brunt--V%
\"{a}is\"{a}l\"{a} frequency, $H_{\mathrm{r}}$ is the total thickness of the
active fluid layer, and $g_{\mathrm{r}}$ denotes the vertically averaged
buoyancy. All these three quantities are held constant. The reference
buoyancy then varies linearly from $g_{\mathrm{r}}(1\mp S)$ at the top of
the active layer to $g_{\mathrm{r}}(1\pm S)$ at the base of the active
layer. Physically acceptable values of $S$ must be such that $0<S<1$ \cite%
{Ripa-JFM-95,Beron-Ripa-97}. Figure \ref{fig}a compares, as a function of $%
S, $ the phase speed of long gravity waves assuming exact dynamics and (\ref%
{IL1n}) with $n=1$, 2, and 3. The comparison is perfect for the barotropic
mode phase speed even including only one layer. The use of two layers
amounts to an excellent representation of the first baroclinic mode phase
speed, and a very good representation of the second baroclinic mode phase
speed. To reasonably represent also the third baroclinic mode phase speed,
no more than three layers are needed.

We now turn our attention to classical baroclinic instability. We thus
consider a \textit{basic state} with a parallel current in an infinite
channel on the $f$ plane, which has a uniform vertical shear and that is in
thermal-wind balance with the across-channel buoyancy gradient. We further
set the basic velocity to vary (linearly) from $2U$ at the top of the active
layer to $0$ at the base of the active layer. Accordingly, the basic
buoyancy field varies from $g_{\mathrm{r}}(1-2fUy/H_{\mathrm{r}}\mp S)$ at
the top of the active layer to $g_{\mathrm{r}}(1-2fUy/H_{\mathrm{r}}\pm S)$
at the base of the active layer ($y$ denotes the accross-channel
coordinate). Figure \ref{fig}b compares in the classical Eady limit, as a
function of the along-channel wavenumber the growth rate%
) of the most unstable normal-mode perturbation assuming exact dynamics and (%
\ref{IL1n}) with $n=1$, 2, and 3. The comparison is almost perfect when
three layers are included and very good when two layers are included. When
only one layer is considered, the comparison is not as good but a high
wavenumber cutoff of baroclinic instability is present. This important
dynamical feature cannot be represented with a slab single-layer model.

\section{Preliminary Conclusions\label{conclusions}}
\vspace{-.1in}

We have tested the performance of a novel inhomogeneous-density
primitive-equation layered model, which features vertical shear
and stratification within each layer, in two important aspects of
ocean dynamics, namely linear waves and baroclinic instability.
Preliminary results suggest that a model with a small number of
layers may be used as a basis for a quite accurate and numerically
economic ocean model. To make stronger statements on the model's
accuracy and efficiency, fully nonlinear, forced--dissipative
problems must of course be considered. This work is currently
underway.

\end{document}